\documentstyle[12pt]{article}
\begin{document}
\pagenumbering{arabic}
{\bf \quad} \\ [0.5cm]
\begin{flushright}
{\bf ISU-IAP.Th93-03,\ Irkutsk}
\end{flushright}
\ \\[1.5cm]
\begin{center}
{\Large \bf $\pi^+$ and $\pi^0$ polarizabilities from
{$\gamma\gamma\rightarrow\pi\pi$} data  \\[0.2cm]
 on the base of S-matrix approach
.}\\[1.5cm]
{\Large A.E.Kaloshin and V.V.Serebryakov${}^1$ }
\end{center}

\begin{center}
\sl Institute of Applied Physics, Irkutsk State University,\\
664003, Irkutsk, Russia.\ \ kaloshin@physdep.irkutsk.su \\
${}^1$ Institute for Mathematics, 630090, Novosibirsk, Russia.\\
\end{center}
\ \\[0.5cm]
\begin{center}
{\Large Abstract.}
\end{center}

We suggest the most model-independent and simple description of the
$\gamma\gamma\rightarrow\pi\pi$ process near threshold in framework of
S-matrix approach. The amplitudes contain the pion polarizabilities
and rather restricted information about $\pi \pi$ interaction. Application
of these formulae for description of
MARK-II \cite{M2}  and Crystal Ball \cite{CB} data gives:
$(\alpha-\beta)^{C}=(6.0\pm 1.2)\cdot 10^{-42} {\rm cm}^{3}$,
$(\alpha-\beta)^{N}=(-1.4\pm 2.1)\cdot 10^{-42} cm^3$ (in units system
$e^2 = 4 \pi \alpha$)
at the experimental
values of $\pi \pi$ scattering lengths. Both values are compartible with
current algebra predictions.
\newpage
\begin{center}
{\bf 1.Introduction.}
\end{center}

The appearing data on
{$\gamma\gamma\rightarrow\pi^{+}\pi^{-}$} MARK-II \cite{M2} and
{$\gamma\gamma\rightarrow\pi^{0}\pi^{0}$} Crystal Ball \cite{CB}
near the threshold give the opportunity to obtain the pion
polarizability from $\gamma \gamma$ experiments. Recall that for
the $\pi^{+}$ polarizability there exist two measurements in processes
on nuclei \cite{Antipov,Aibergenov} but for $\pi^{0}$ the $\gamma \gamma$
experiments give the first possibility to measure it. The polarizability's
measurements allow to verify the predictions of different low-energy
models based on chiral symmetry. Note that a straightforvard using of
chiral model for description of $\gamma\gamma\rightarrow\pi\pi$
cross--section does not have big sense becouse of huge unitary effects.
One can see (see, for instance, \cite{Penn}) that the one--loop chiral
model calculations \cite{chiral} differ significally from unitary
formulae and experiment even in region of 400--500 Mev.

By this reason we use our S-matrix approach \cite{KS-86} to this reaction.
Note that in other unitary models \cite{Menn.,Lyth,MP} polarizabilities
are not controlled and there exist some problems with low-energy theorem
even on classical level.

The polatizabilities in the $\gamma\gamma\rightarrow\pi\pi$ amplitude
play the same role as the electromagnetic radius in a formfactor:
they are the structure constants in threshold decomposition.
So the problems at extraction of $\alpha, \beta$ from a cross--section
are in principal the same as in $<r_{\pi}^2>$ extraction from pion
formfactor at $q^2> 4m^2_{\pi}$. The main problem is that the
polarizability's effects can not be separated from effects of $\pi\pi$
--interaction in final state and one is forced to utilize this information
in the $\gamma\gamma\rightarrow\pi\pi$ amplitude. Note that the pion
polarizabilities can be obtained from any analysis of
$\gamma\gamma\rightarrow\pi\pi$ data in wide energy region by extrapolation
to point $s=0, t=m_\pi^2$. Such an analysis is in fact rather complicated
task (in our opinion it is not made correctly up to now) which needs
first of all the better understanding of $I=J=0$ hadron dynamics.
So we think it will be reasonable to develop the more model--independent
and simple description of near-threshold region
$\gamma\gamma\rightarrow\pi\pi$ which includes only some well--established
properties of $\pi\pi$--interaction. We hope that future
$\gamma\gamma\rightarrow\pi\pi$ experiments can utilize these formulae
in physical analysis of data.

In \cite{KS-91} we have been performed the simplest analysis of
{$\gamma\gamma\rightarrow\pi^{0}\pi^{0}$} \cite{CB} data to extract
$(\alpha-\beta)^{\pi^{0}}$. Here we analyse the obtained formulae and
use them for description of data \cite{M2,CB} on both reactions.
\newpage

\begin{center}
{\bf 2. Simplified unitary formulae for
$\gamma\gamma\rightarrow\pi\pi$ near the threshold.}
\end{center}

Let us introduce the helicity amplitudes $T_{\Lambda}$ in c.m.s. in
usial manner. The cross--section $\gamma\gamma\rightarrow\pi^{+}\pi^{-}$
for unpolarized photons is defined as
\begin{equation}
\frac{d \sigma}{d \Omega} = \frac{\rho(s)}{2s (8 \pi)^2}
\{{\mid T_{++}\mid}^2 + {\mid T_{+-}\mid}^2 \}
\end{equation}
where $\rho(s) = (1 - 4\mu^2/s)^{1/2}$ , $\mu = m_\pi$.
For $\gamma\gamma\rightarrow\pi^{0}\pi^{0}$  one should add the factor
$ 1/2! $ in (1) to take into account the identity of $\pi^0$'s.
Isotopical amplitudes $\gamma\gamma\rightarrow\pi\pi$ are:
\footnote{\small We indicate these relations here to avoid the mistake as in
some
recent papers \cite{MP,Harjes} on this matter, where the $T^N$ in (2) has the
opposite sign. The reason of this mistake may be found in any old
textbook: you should denote $\mid \pi^+> = - \mid I=1,I_3=1> $
to use the standard Clebsh-Gordan coefficients.}
\begin{eqnarray}
T^C = T(\gamma\gamma\rightarrow\pi^{+}\pi^{-}) = \sqrt{\frac{1}{3}}\ T^{I=0}
 + \sqrt{\frac{1}{6}}\ T^{I=2}  \nonumber \\
T^N= T(\gamma\gamma\rightarrow\pi^{0}\pi^{0}) = \sqrt{\frac{1}{3}}\ T^{I=0}
 - \sqrt{\frac{2}{3}} \ T^{I=2}
\end{eqnarray}
It is well known that helicity amplitudes contain the kinematical
singularities and zeroes. For this reaction we can pass over to reduced
amplitudes $M_{\Lambda}$ which do not contain them \cite{AG}.
\begin{equation}
T_{++} = s M_{++} ,\ \ \ T_{+-} = (tu - \mu^4) M_{+-}
\end{equation}
Besides the using of amplitudes $M_{\Lambda}$ is very useful in another
aspect: it allows to control easy the low-energy theorem including
the structural corrections -- see below. So we shall use the reduced
amplitudes $M_{\Lambda}$ in the following.

The starting "final state interaction" formulae for $M_{++}$ are contained
in \cite{KS-86}. They satisfy the one-channel unitary condition for
lowest partial wave and are normalized on the pion polarizability.
Note that the polarizability is the attribute of just the lowest
partial wave. The formulae of \cite{KS-86} have two arbitrary constants
$C^0$ and $C^2$ , if the contributions of nearest cross--exchanges
are known. But one can refuse from this additional suggestion by introducing
of some effective constants which absorb any high--energy contributions.
Only the lightest $\rho,\omega$ exchanges should be considered as "alive".
As a result we have the following expressions for reduced helicity
amplitudes:
\begin{eqnarray}
{\bf M}^{C}_{++}&=&\frac{{\Omega}^{0}(s)}{\sqrt{3}}(C^{0}+
H^{0}_{V}(s))+\frac{{\Omega}^{2}(s)}{\sqrt{6}}(C^{2}+H^{2}_{V}(s))
+{\bf W}^{C}(s,t) \nonumber \\
{\bf M}^{N}_{++}&=&\frac{{\Omega}^{0}(s)}{\sqrt{3}}(C^{0}+
H^{0}_{V}(s))-\sqrt{\frac{2}{3}}{\Omega}^{2}(s)(C^{2}+H^{2}_{V}(s))
+{\bf W}^{N}(s,t) \ \
\end{eqnarray}
Here $\Omega^{I}$ is the Omnes function, corresponding to phase shift
$\delta^{I}_{0}$.
\begin{eqnarray}
\Omega^{I}(s) &=& exp\{\ \frac{s}{\pi}\int \frac{ds'}{s'(s'-s)} \delta^{I}_{0}
(s')\ \} \nonumber  \\
{\bf W}^{C}(s,t) &=& W^{\pi}(s,t)+W^{\rho}(s,t)+a^{C}\\
{\bf W}^{N}(s,t) &=& W^{\rho}(s,t)+W^{\omega}(s,t)+a^{N} \nonumber
\end{eqnarray}
\[
{\bf W}^{\pi}=\frac{2e^{2}\mu^{2}}
{(t-\mu^{2})(u-\mu^{2})}, \quad
{\bf W}^{\rho}= -Z^{\rho}(\frac{t}{m^{2}_{\rho}-t}-\frac{u}
{m^{2}_{\rho}-u}), \quad Z^{R} = g^{2}_{R\pi\gamma}
\]
Functions $H^{I}_{V}(s)$ in (4) are the rescattering contributions:
\begin{equation}
{\bf H^{I}_{V}(s)}=\frac{s}{\pi}\int \frac{ds'}{s'(s'-s)} V^{I}(s')
\frac{sin \delta^{I}}{\mid\Omega^{I}\mid}
\end{equation}
$a^{C}$,$a^{N}$ in (5) are some effective constants from cross--contributions
and possibly s--channel high energy contributions. If the decays widths
$R\rightarrow\pi\gamma$ are known, they may be estimated (see below)
but we prefer to consider them as free parameters.

The J=0 projection of (4) must satisfy the one-channel unitary condition
for isotopical amplitudes I=0,2 .
\begin{equation} Im F^{I}(s)=\rho(s)F^{I}(s) f^{*I}_
{\pi\pi\rightarrow\pi\pi},\ \ \rho(s)=(1-4\mu^2/s)^{1/2}
\end{equation}
It is easy to see that the unitary condition is reduced to
\begin{equation} V^{I}(s) = W^{I,J=0}(s)
\end{equation}
where  $W^{I}$ are the isotopical combinations of $W^{C},W^{N}$ (5)
which are projected on J=0 state.

The amlitudes (4) contain four free parameters:
$C^{0},C^{2}, a^{C},a^{N}$. But becouse of smallness of $\delta^{2}_{0}$
phase shift the Omnes function $\Omega^{2}$ is very close to 1 and
$C^2 \Omega^{2}$ does not differ from constant . So we can put $C^{2}=0$
in (4). As a result we have 3 free parameters for two reactions
$\gamma\gamma\rightarrow\pi^{+}\pi^{+},\pi^{0}\pi^{0} $. The pion
polarizabilities are related with them as
\begin{eqnarray} &&\frac{\mu}{2}
(\alpha-\beta)^{C} = \frac{C^{0}}{\sqrt{3}}  +  a^{C}- Z^{\rho}
\frac{2\mu^{2}}{m^{2}_{\rho}-\mu^{2}} \nonumber\\ &&\frac{\mu}{2}
(\alpha-\beta)^{N} = \frac{C^{0}}{\sqrt{3}}  +  a^{N}- Z^{\rho}
\frac{2\mu^{2}}{m^{2}_{\rho}-\mu^{2}}- Z^{\omega}
\frac{2\mu^{2}}{m^{2}_{\omega}-\mu^{2}} \end{eqnarray}

In the folllowing we use  $(\alpha-\beta)^{C},
(\alpha-\beta)^{N}, C^{0}$ set of parameters. The reason of appearing
the additional parameter $C^{0}$ is clear: the point of polarizability's
definition is out of physical region of $\gamma\gamma\rightarrow\pi\pi$
and final state interaction leads to  $\Omega^{0}(s)\not= \Omega^{0}(0)$.

Let us note that amplitudes (4) may be rewritten in another equavalent form.
If $\delta^{0}_{0}\rightarrow\pi$  at  $s\rightarrow \infty$ , there exists
the equality
\begin{equation}
\frac{1}{\Omega^{0}(s)} = 1+\beta^{0}s+\frac{s}{\pi}
\int \frac{ds'}{s'(s'-s)} Im\frac{1}{\Omega^{0}(s')},\nonumber
\end{equation}
and similar for I=2 without $\beta s$ term. It leads to disappearing
of constants $a^C,\ a^N$ in (4), to appearing of $C^2$ and replacement
of $C^0$ by polinom. That is an equavalent form, but we prefer to use
the form (4).
\begin{center}
{\bf 3. About the $\pi\pi-$ interaction.}
\end{center}

Recall \cite{KS-86} that the formulae (4) satisfy the one--channel
unitary condition for lowest partial wave J=0 and $ \delta^{0}_{0}$
denotes only smooth part of experimental phase shift without the sharp
$f_{0}(975)$ effect. For justification of such approach one should
consider the multichannel problem (see, for instance, \cite{KS-88})
but it can be easy understood at qualitative level. Simplest (and widely
used) suggestion on the structure of $\pi\pi$ amplitude is (see review
\cite{ADS}):
\begin{equation}
\delta^{0}_{0} = \delta^{Backg} +\delta^{f_{0}} ,\qquad
      \eta^{0}_{0} = \eta^{f_{0}}
\end{equation}
In any case the $\pi\pi\rightarrow\pi\pi$  J=I=0 amplitude contains
the smooth contribution with typical scale $\sim 500$ Mev and the sharp
$f_{0}$ one with scale $\sim 50$ Mev. Solving the multichannel problem
\cite{KS-88} we see that these two contributions are separated from
each other in the answer. If the coupling $f_{0}\gamma\gamma$ is very
small ( that is
an experimental fact, the decay width $f_{0}\rightarrow\gamma\gamma$
is the value $ \sim 0.1 \ Kev $ ) then the presence of $f_{0}$ leads only
to the local interference effects near 1 Gev \cite{KS-88} and its influence
is negligible in a big energy scale . So for region  $\sqrt{s} < 800 \ Mev $
the effective one--channel problem is arised and as a result we have
the amplitudes (4).

So to calculate the $\gamma\gamma\rightarrow\pi\pi$ we should use
the main properties of smooth part $\delta^{0}_{0}$ phase shift.
Firstly that is the threshold region which is described in term of
scattering lenghs and the chiral zeroes.
\begin{equation}
tg \delta^{I}_{0} = a^{I}_{0} \mu \rho(s) \frac{(s-s^{I}_{0})}
{(s+s_{1})}\frac{(4\mu^{2}+s_{1})}{(4\mu^{2}-s^{I}_{0})} ,
\end{equation}
where $s_{1} \sim m^{2}_{\rho}$ . This formula was used for describing
the $\pi\pi$ data and the analysis \cite{Alekseeva} gives:
\begin{eqnarray}
&&s^{0}_{0} = -0.2 \mu^{2},\qquad s^{2}_{0} = 2.4 \mu^{2}, \nonumber\\
&&a^{0}_{0}\mu = 0.24\pm 0.04,\quad a^{2}_{0}\mu =- 0.04\pm 0.04
\end{eqnarray}
Secondly, we slould utilise the large scale behaviour of phase shift.
In experiment smooth part of $\delta^{0}_{0}$ crosses the $\pi/2$
at $\sqrt{s} \sim 900$ Mev and looks as linear on $\sqrt{s}$ function in
very wide energy region.

In this situation we describe threshold region as (12), then the phase shift
is described as linear on $\sqrt{s}$ function, and it reachs $\pi$ at
$\sqrt{s}$ = 1.5 Gev---see Fig. 1,2.
Another variant of action may consist in using for amplitude
 $\pi\pi\rightarrow\pi\pi$  the  N/D represantation type of \cite{VDS}
which includes all above--mentioned properties. We can also modify
sligtly this model for better describing of experiment. But we have been
convinced that the phase shift behaviour at say 1.3 Gev practically
does not influence on the extracted values of polarizabilities. So
we use below the simplest linear approximatiom with threshold behaviour
(12). As for the phase shift $\delta^{2}_{0}$, we need only the threshold
properties (12) becouse it influences only on the rescattering integral
which is saturated by the threshold region.
\begin{center}
{\bf 4.  The low--energy theorem and polarizabilities.} \end{center}

It is well known (see, for instance, review \cite{Petrun'kin}) that
Compton scattering amplitude on hadron contains the classical charge
normalization \cite{Low} and next structural terms of threshold
decomposition. In case of spin 0 target there appear in $\omega^2$ order
two structural constants $\alpha,\beta$, named the electriical and
magnetic polarizabilities. One can relate them with the helicity
$\gamma\gamma\rightarrow\pi\pi$ amplitudes \cite{KS-86}, which have the
following form in the vicinity of point  $s=0,t=\mu^2$ :
\begin{eqnarray}
&&T_{++}=s\ [\ M_{++}^{\pi} +\frac{\mu}{2}(\alpha-\beta)^{\pi}\ ] \nonumber\\
&&T_{+-}=(tu-\mu^4)\ [\ M_{+-}^{\pi} +\frac{(\alpha-\beta)^{\pi}}{2\mu}\ ]
\end{eqnarray}
Here $M^{\pi}$ is the contribution of $\pi$--exchange (QED) for the
charged pions and zero for neutral. Note that here the reduced helicity
amplitudes m \cite{AG} are arised , which are useful to control the
low--energy theorem.

Up to now there exist two measurements of $\pi^+$ polarizability in
experiments on a target. The experiment
{$\pi Z \rightarrow\pi\gamma Z$} \cite{Antipov} gives the value
$(\alpha-\beta)^{C}=(17.0\pm3.6)\cdot 10^{-42} cm^{3}$, the result of
another experiment
{$\gamma p\rightarrow \gamma \pi^{+} n $} \cite{Aibergenov}
is $(50\pm30)\cdot 10^{-42} cm^{3}$.

As for different model prediction for $\alpha,\beta$, let us begin from
current algebra (see review \cite{Terent'ev} on this matter). The using of
PCAC leads to $(\alpha-\beta)^{\pi^0} = 0$,
the polarizability of {$\pi^{+}$} is reduced to formfactor $\pi V A$,
which is measured in the decay $\pi\rightarrow\gamma e \nu$.
The using of last experimental data for this decay \cite{Decay_pi} leads to:
\begin{equation}
(\alpha-\beta)^C =( 8 \div 10)\cdot 10^{-42}\ cm^3 \nonumber
\end{equation}
The majority of model predictions, based on different forms
of the chiral model, also are near this value or slightly higher.

As for neutral pion polarizability, all models predict smaller
( or much smaller) value as compared with charged pion. But the value
( and sign) depends strongly on the model. For instance in the quark--
virton model \cite{Dineihan} $(\alpha-\beta)^N = 2.8\cdot 10^{-42}\ cm^3 $ ,
in the four--quark model of superconductor type\cite{Volkov}
 $(\alpha-\beta)^N = - 3.2 \cdot 10^{-42}\ cm^3 $.
\begin{center}
{\bf 5. Our preliminary expectations.}
\end{center}

In framework of the approach \cite{KS-86} one can obtain some estimates
of the parameters in (4). The starting point of \cite{KS-86} is the
dispersion relation for the reduced helicity amplitude $M_{++}$ at
fixed $t\sim\mu^{2}$, which is
written without subtractions according to the standard Regge behaviour.
Putting  s=0 in it and saturating the cross--integral by the nearest
resonance exchanges we shall obtain
\begin{eqnarray}
&&\frac{\mu}{2} (\alpha-\beta)^{C} =
-Z^{\rho}+Z^{b_{1}}+Z^{a_{1}}+ \frac{5}{3} Z^{a_{2}}
+ \frac{C^{0}}{\sqrt{3}} + \frac{C^{2}}{\sqrt{6}} \nonumber\\
&&\frac{\mu}{2}(\alpha-\beta)^{N} =
-Z^{\rho}-Z^{\omega}+Z^{h_{1}} +Z^{b_{1}}+ \frac{C^{0}}{\sqrt{3}}
- \sqrt{\frac{2}{3}} C^{2}
\end{eqnarray}
Here $C^{0},C^{2}$ are the corresponding s--channel inegrals at s = 0.
If to put  $C^{2}=0$ and to subtract one equation from another
there arises the relation between polarizabilities of charged and
neutral pions. But in fact in I=2 channel there exists one contribution,
which can not be neglected --- that is the rescattering. But this
contribution can be easy calculated becouse it is defined only by the
threshold $\pi \pi$ properties. So taking the experimental values (13)
we shall obtain $C^{2}=-0.0328 \ Gev^{-2}$ and the relation between
polarizabilities has the form:
\begin{equation} (\alpha-\beta)^{C}-(\alpha-\beta)^{N} =
9.1\cdot 10^{-42}\quad cm^{3}\quad   ( 12.6\quad at \quad C^{2}=0)
\end{equation}
if to take the existing values \cite{PDG} for $R\rightarrow\pi\gamma$.
\footnote{\small The relation (17) demonstrates one of the possible
contradictions
with using the wrong isotopics \cite{MP,Harjes}. It leads roughly to
the changing of sign in (17) and contradicts evidently to present
experimental and theoretical knowledge about $\alpha,\beta$.}
Note that the scale and sign of right side (17) is defined in fact
by the decay width $\omega\rightarrow\pi\gamma$ which is measured
with best accuracy. As a result we can estimate
$(\alpha-\beta)^{C}, C^{0}$:
\begin{eqnarray}
(\alpha-\beta)^{C}=9\cdot 10^{-42}\quad cm^{3},\quad C^{0}=0.106\
Gev^{-2},\nonumber\\
at \quad (\alpha-\beta)^{N} << (\alpha-\beta)^{C}
\end{eqnarray}
We can expect such scale values for $(\alpha-\beta)^{C}, C^{0}$
in case of absence of different exotic phenomena ( big odderon
contribuion in $\gamma\gamma\rightarrow\pi^{0}\pi^{0}$ \cite{Zh},
some unknown dynamics in I=2 channel,...). Let us note also that
the smallnes of $\pi^{0}$ polarizability as it seen from (16)
is the result of cancellation of rather big different contributions
near the point $s=0,t=\mu^{2}$. So it does not means the smallness of
$\gamma\gamma\rightarrow\pi^{0}\pi^{0}$ amplitude in
wide kinematical region.
\begin{center}
{\bf 6. The analysis of $\gamma\gamma\rightarrow\pi\pi$ data.}
\end{center}

The numerical analysis of the obtained amplitudes shows that the most
favourable region for extraction of $\alpha,\beta$
is  $\sqrt{s} < 500 \ Mev$, where practically closed description
is arised, i.e. here the function $\Omega^{0}(s)$ is defined mainly
by the threshold properties of phase shift. Besides the contributions
of $\rho,\omega$ exchanges is not essential here. So let us start from
consideration of this region.
\begin{center}
{\bf 6.1 Near--threshold region: $\sqrt{s} < 500 \ Mev $ .}
\end{center}

In this region there exist the MARK-II data \cite{M2}
\footnote{\small In the region  $\sim$ \ 400 Mev in the process
$\gamma\gamma\rightarrow\pi^{+}\pi^{-}$ the helicity amplitude
$M_{+-}$ is not negligible. But here it practically coinside with QED
contribution \cite{KS-86} , so we take into  account only this contribution.
At higher energy it is nesessary to construct the model for
helicity 2 amplitude too.}
( 5 points between 350 and 400 Mev ) and Crystal Ball \cite{CB}.
The quality of data does not allow to perform the total 3-parameter
analysis, so let us put $(\alpha-\beta)^{N} = 0$. As we shall see later
in all variants of analysis the neutral pion polarizability is
compatible with zero.
Besides it was found that the cross--section
$\gamma\gamma\rightarrow\pi^{0}\pi^{0}$ depends essentially on all
three parameters whereas $\gamma\gamma\rightarrow\pi^{+}\pi^{-}$
below 400 Mev only weakly depends on the $(\alpha-\beta)^{N}$.
Unlike of \cite{KS-91} we use exact value of $m_{\pi^{0}}$ in phase
volume which is essential near threshold.
\footnote{\small In amplitudes all masses are equal to $m_{\pi^{+}}$.
In principal the effects of mass difference could play some role in
amplitudes too becouse of existence of QED "amplifier" in rescattering.
We checked it on the base of K-matrix approach and found that amplitudes
are sensitive to the exact value of $\pi^{+}$ mass but not to
$m_{\pi^{0}}$. This fact is defined by the unitary constrains and by
presence of chiral zeroes in $\pi \pi$ amplitudes.}

1). Let us try to describe MARK-II \cite{M2} data separately. We shall
obtain very broad distributions for $\chi^{2}$, as for $C^0$
here is practically a plateau. It means that cross--section is
sensitive only to one effective parameter --- some linear combination
of $(\alpha-\beta)^{C}$ and $C^0$. This fact may be seen from Fig.3
where the lines (almost linear) of equal $\chi^{2}$ are depicted
in plane of parameters. Any point on this line gives the same
cross--section in region of MARK-II threshold data but at higher energies
they are very different.

2). Describe only Crystal Ball \cite{CB} data in this region.
Again very broad distributions are arised, the minimum of $\chi^{2}$
corresponds to $(\alpha-\beta)^{C} \sim 5\cdot 10^{-42} cm^{3}$.
One can see however that the data \cite{M2,CB} do not contradict to each
other.

3).Describe the data \cite{M2} and  \cite{CB} together. It leads to:
\begin{equation}
(\alpha-\beta)^{C}=5.4\cdot 10^{-42} cm^{3},\qquad
\chi^{2}=15.4 \quad at \quad NDF=8
\end{equation}
The $C^0$ parameter is defind very badly -- situation is similar to
Fif.3.

\begin{center}
{\bf 6.2  Fit in extended region.}
\end{center}

Let us extend the energy region of Crystal Ball data used in analysis
up to 850 Mev. At higher energies helicity 2 amplitude becomes
essential \cite{KS-86} but variation of this bound does not change
the results essentially.

1). Fit only CB data by three parameters.

As a result we have been found the existence of two minimums with close
values of $\chi^{2}$ :
\begin{eqnarray}
I.\ \ \ \ (\alpha-\beta)^{C}=7.0\cdot 10^{-42} cm^{3},\
(\alpha-\beta)^{N}=-0.15\cdot 10^{-42} cm^3,\nonumber \\
 C^{0}=0.156\ Gev^{-2},\ \
\chi^{2}=9.1\ \ at \ \ NDF=9 \ \ \ \ \ \ \ \ \ \ \nonumber\\[0.5cm]
II.\ \ \ \ (\alpha-\beta)^{C}=8.5\cdot 10^{-42} cm^{3},\
(\alpha-\beta)^{N}=0.4\cdot 10^{-42} cm^3, \ \ \nonumber \\
C^{0}=0.026\ Gev^{-2}, \ \
\chi^{2}=9.4\quad at \quad NDF=9  \ \ \ \ \ \ \ \
\end{eqnarray}

The second minimum leads to sharp increasing of both cross--sections
above the region of fit which contradict to CB data here. So we shall
not consider this variant in the following. It is interesting that both
mimimums do not contradict to theshold MARK-II data --- see Fig.3.

2). Fit the Crystal Ball ( $\sqrt{s} < 850 \ Mev$) and MARK-II
($\sqrt{s} < 400 \ Mev$) data together.

Two-parameter fit,$(\alpha-\beta)^{N}=0 $, gives:
\begin{eqnarray}
(\alpha-\beta)^{C}&=&(6.1\pm 1.2)\cdot 10^{-42} cm^{3},\ \
C^{0}=0.160\pm 0.012\ Gev^{-2},\ \nonumber  \\
& & \chi^{2} = 18.4\ \ at \ \ NDF=15
\end{eqnarray}

Three-parameter fit:
\begin{eqnarray}
(\alpha-\beta)^{C}&=&(6.0\pm 1.2)\cdot 10^{-42} cm^{3},\ \
(\alpha-\beta)^{N}=(-1.4\pm 2.1) \cdot 10^{-42} cm^3, \nonumber \\
& & C^{0} = 0.150\pm 0.020\ Gev^{-2},\nonumber \\
& & \chi^{2} = 17.9\ \ at \ \ NDF=14
\end{eqnarray}

The corresponding cross--sections are depicted at Fig.4,5,6.

\begin{center}
{\bf 7. Conclusion.}
\end{center}

So we tried in the first time to obtain an information about
pion polarizabilities from the existing data on
$\gamma\gamma\rightarrow\pi\pi$. Here we delibertely used the most
simple description of amplitudes on the base of S-matrix approach.
Such an approach may be called as "structureless" since it utilizes
only some properties of the smooth $\pi \pi$ amplitude without
consideration of its physical structure.
We suppose that for region of $\sqrt{s} < 800 \ Mev$ such description
is fully adequate and allows to extract pion polarizabilities from
experiment. We think that more detailed investigation of this question
with consideration of multichannel problem ( and with correction of some
defects in existing models) will be useful. Firstly, it will allow
to define more precisely the region of application of simplified
formulae and to obtain more correct values of polarizabilities.
Secondly, it is interesting to analyse the spectrum and properties
of scalar mesons from joint consideration of hadron and two--photon
experiments. Let us note also very unexpected results (from point of
view of threshold characteriscs first of all) of CELLO group
\cite{Harjes} on partial-wave analysis of
$\gamma\gamma\rightarrow\pi^{+}\pi^{-}$ above 0.8 Gev.

As a result of our analysis of Crystal Ball and MARK-II data we can
note the following facts:
\begin{itemize}
\item The data of CB \cite{CB} and MARK-II \cite{M2} are described well by
formulae (4) at the standard form of $\pi\pi$ interaction and do not
contradict to each other.
\item The data of CB and MARK-II separately indicate on rather small value
of $\pi^{+}$ polarizability which is compartible with current algebra
but contradicts to experimental value $(17.0\pm3.6)\cdot 10^{-42} cm^{3}$ from
{$\pi Z \rightarrow\pi\gamma Z$} \cite{Antipov}. The value of $\sim 12.0$
is excluded by the $\gamma\gamma\rightarrow\pi\pi$ data.
\item At any variant of analysis the neutral pion polarizability
is compartible with zero in accordance to current algebra.
\item At changing of scattering length $a^{0}_{0}=0.24$ to any side
the disagreement of Crystal Ball and MARK-II data is arised.
\item The obtained values of $\alpha-\beta$ at least qualitatevely
correspond to our estimates (18), based on the saturation by nearest
resonances. So we do not see some new effects in such analysis.
\end{itemize}

When this work was finished we received the paper \cite{Penn}, where
an ettempt was made to extract $\pi^0$ polarizability from data
on the base of multichannel model \cite{MP}. Note that Pennington
obtains  $(\alpha - \beta)^N > 0 $ with much less errors than ours.
But one should note few defects in model \cite{MP} including the
wrong isotopics.

\newpage
\begin{center}
{\bf Figure captions.}
\end{center}
\begin{itemize}
\item {\bf Fif.1} The phase shifts described it text at the threshold
parameters (13).
\item {\bf Fig.2} The Omnes functions corresponding to Fig.1.
\item {\bf Fig.3} Lines of equal $\chi^2$ at fiting of 5 threshold
points of MARK-II by two parameters, $(\alpha-\beta)^{N}=0 $ .
Different scattering lengths $a^0_0$ are indicated.  The error (one standard
deviation ) coinsides approximately with distance between lines.
Along these lines $\chi^2$ is the same: $\chi^2=8.4$.
\item {\bf Fig.4}
Comparison of MARK-II data with best fit curve.  All curves with equal
$\chi^2=8.4$ coinside in this region but differ at higher energies
$\sqrt{s} = 700 - 800 \ Mev$. The helicity 2 QED contribution is
subtracted from MARK-II data.
\item {\bf Fig.5}
The cross--section $\gamma\gamma\rightarrow\pi^+\pi^-$ of helicity 0
at higher energies . Two variants correspond to two sets of parameters (20).
\item {\bf Fig.6} The cross--section $\gamma\gamma\rightarrow\pi^0\pi^0$.
Comparison of CB data with best fit curve at values (20).
\end{itemize}
\end{document}